\documentclass[12pt]{article}
\usepackage{cite,epsfig,amssymb,amsmath,graphicx,color}
\newcommand{\nn}{\nonumber}

\begin{document}

\vspace{9mm}

\begin{center}
{{{\Large \bf Abelian Projections of the Mass-deformed ABJM Theory and Weakly Curved Dual Geometry}
}\\[17mm]
Young-Hwan Hyun$^{1}$,~~Yoonbai Kim$^{1}$,~~O-Kab Kwon$^{1,3}$,~~D.~D. Tolla$^{1,2}$\\[3mm]
{\it $^{1}$Department of Physics,~BK21 Physics Research Division,
~Institute of Basic Science, Sungkyunkwan University, Suwon 440-746, South Korea\\
$^{2}$University College,\\
Sungkyunkwan University, Suwon 440-746, South Korea}\\[2mm]
{\it $^{3}$Institute for the Early Universe, Ewha Womans University, Seoul 120-750, South Korea}\\[2mm]
{\tt yhhyun@skku.edu,~yoonbai@skku.edu,~okab@skku.edu,~ddtolla@skku.edu} }
\end{center}
\vspace{20mm}

\begin{abstract}
We construct ${\cal N}=2,4$ supersymmetric Abelian projections of the
${\cal N}=6$ mass-deformed Aharony-Bergman-Jafferis-Maldacena (ABJM) theory. There are well-defined dual background geometries for the ${\cal N}=2$ Abelian theory, while those geometries are unclear for the ${\cal N}=4$ Abelian theory. The ${\cal N}=2$ theory is built on the supersymmetric vacua of the mass-deformed ABJM theory, which are proven to have a one-to-one correspondence with the ${\mathbb Z}_k$ quotient of Lin-Lunin-Maldacena geometries.
We select one special vacuum of the mass-deformed ABJM theory and show that the corresponding
geometry is weakly curved at every point of the entire space transverse to the M2-branes
in the large-$N$ limit.
\end{abstract}

\newpage
\tableofcontents

\section{Introduction}

The AdS/CFT duality is a powerful tool for studying strongly coupled gauge
theories~\cite{Maldacena:1997re,Gubser:1998bc,Witten:1998qj}.
This duality has been tested in numerous examples in a string-/M-theory framework and
applied to various physical models.
In particular, there is accumulating evidence that the AdS/CFT duality
can be used to study physics in the strongly coupled regime in condensed
matter systems~\cite{Hartnoll:2009sz,Herzog:2009xv,McGreevy:2009xe}.

An important example of the AdS/CFT duality related to a low-energy effective
action of $N$ coincident M2-branes on the ${\rm {\mathbb C}}^4/{\mathbb Z}_k$ orbifold fixed point
was proposed in Ref. \cite{Aharony:2008ug}.
This effective theory is called Aharony-Bergman-Jafferis-Maldacena (ABJM) theory and
was conjectured to be dual to type IIA string theory
on ${\rm AdS}_4\times \mathbb{CP}^3$
or M-theory on ${\rm AdS}_4\times {\rm S}^7/\mathbb{Z}_k$, according to the value of
the Chern-Simons (CS) level $k$.
Since the boundary of the AdS$_4$ space is conformally mapped to
the flat $(2+1)$-dimensional Minkowski space,
the dual field theories are closely related to the strongly coupled theories
of planar condensed matter systems. The direct application of ABJM theory to specific condensed matter systems
seems far from possible, due to its complicated non-Abelian structure
and huge number of degrees of freedom.

In order to make more realistic attempt,
a consistent Abelian truncation of the bosonic part of the
${\cal N}=6$ mass-deformed ABJM (mABJM) theory~\cite{Hosomichi:2008jb,Gomis:2008vc} is probably a promising direction~\cite{Mohammed:2012gi}.
In the construction of an Abelian theory, an ansatz is used in terms of some constant
matrices, called the Gomis-Rodriguez-Gomez-Van Raamsdonk-Verlinde (GRVV) matrices~\cite{Gomis:2008vc}.
These matrices were originally introduced to obtain discrete vacua of the
mABJM theory.
The Abelian truncation in Ref.\cite{Mohammed:2012gi} preserves
${\cal O}(N)$ degrees of freedom out of ${\cal O}(N^2)$ degrees of freedom of the mABJM theory in which these Abelian fields may describe the collective motions of
${\cal O}(N)$ charged particles.
It was also proposed that the Abelian action governing the collective motions
can be an effective action of a condensed matter system.
Specifically, the relativistic Landau-Ginzburg model was constructed from
the Abelian projected model.
These results suggest that the dual gravity theories of condensed matter systems
can be studied based on the gauge/gravity duality of the mABJM theory.

In this paper, we employ a generalized ansatz
that includes the fermion part of the mABJM theory
to construct a supersymmetric Abelian projected theory.\footnote{See Ref. \cite{Murugan:2013jm} for an alternative approach to obtain a supersymmetric Abelian projected theory.}
First, we show that the resulting theory has ${\cal N}=4$ supersymmetry.
The Abelian theory obtained in Ref.\cite{Mohammed:2012gi} is a special representative
of our construction.
The backgrounds used in the ${\cal N}=4$ Abelian theories are not supersymmetric vacua of the mABJM theory.
As a consequence, there are some problems in finding the gravity dual.
In order to overcome these problems we propose a new ansatz based on the vacua of the mABJM theory,
and construct an ${\cal N}=2$ Abelian CS theory.

Half-Bogomolnyi-Prasad-Sommerfeld (BPS) geometries for M2-branes in the presence of the transverse
4-form flux in 11-dimensional  supergravity are
called the Lin-Lunin-Maldacena (LLM) geometries~\cite{Lin:2004nb,Bena:2004jw}.
These LLM geometries
were expected to be dual to the vacua of the mABJM theory with CS level $k=1$.
However, the number of classical vacua constructed in Ref. \cite{Gomis:2008vc} was more numerous than
the partitions of $N$, which is the number of the half-BPS LLM solutions for a given number $N$ of M2-branes.
This mismatch was fixed by excluding the vacua in which the supersymmetry is broken
dynamically~\cite{Kim:2010mr}.
The one-to-one mapping between the vacua in the mABJM (or mABJ~\cite{Aharony:2008gk}) theory
with CS level $k$ and the LLM geometries with a ${\mathbb Z}_k$ orbifold and discrete
torsions is explained in Ref. \cite{Cheon:2011gv}.
(See also Refs. \cite{Auzzi:2009es,Hashimoto:2011nn}.)

When we consider a gravity dual for a fluctuation on the field theory side,
the corresponding background geometry should
be weakly curved everywhere in the large-$N$ limit.
However, we will show that in general the LLM geometries involve highly curved regions even in the large-$N$ limit.
Therefore, the gravity duals for some fluctuations are not well-defined.

In this paper, we select one special vacuum of the mABJM theory
and confirm that the corresponding dual LLM geometry
is weakly curved everywhere in the large-$N$ limit.
To avoid complication, we focus on the case of $k=1$.
Based on the work of Ref. \cite{Cheon:2011gv}, one can also extend the discussion to the case of $k>1$.
For the LLM geometry, we compute the Ricci scalar at every point of the entire space
transverse to the M2-branes and show that the Ricci scalar has positive values near the boundaries between
the black and white strips in the droplet representation of the LLM geometry.
Furthermore, the Ricci scalar decreases monotonically with increasing distance
from the boundary
and becomes a negative constant asymptotically.
In this asymptotic region the geometry is AdS$_4\times$S$^7$.
Over the entire geometry, the absolute value of the Ricci scalar decreases with increasing $N$.

Some fluctuations of a field theory vacuum with a weakly curved dual
geometry, may lead to well-defined dual
gravity modes, according to the gauge/gravity duality.
For this reason, our ${\cal N}=2$ Abelian theory can be considered
as a special kind of fluctuation on the vacuum and has a well-defined
gravity dual on the corresponding LLM geometry.
Since we consider the duality on the background of M2-branes polarized to M5-branes
in the presence of the 4-form field strength~\cite{Myers:1999ps},
the gauge/gravity which we are considering does not belong to the well-known AdS/CFT duality.
Rather, it is similar to the duality on the background of D3-branes polarized to D5(or NS5)-branes
in type IIB string theory~\cite{Polchinski:2000uf}.

This paper is organized as follows.
In Sec. 2 we summarize the mABJM theory with explicit SU(2)$\times$SU(2)$\times$U(1) global
symmetry and discuss the supersymmetric vacua of the theory.
In Sec. 3 we construct the ${\cal N}=4$ supersymmetric Abelian projection of the mABJM theory with
general ans\"atze in terms of the GRVV matrices and discuss its problems in relation to the gravity dual.
We also construct the ${\cal N}=2$ Abelian theory based on the vacua of the
mABJM theory.
In Sec. 4 we select one special vacuum in the mABJM theory and investigate the corresponding
dual LLM geometry. We conclude in Sec. 5.

\section{mABJM theory and supersymmetric vacua}
The main purpose of this paper is to construct an ${\cal N}=2$ Abelian projected mABJM theory
and to propose a weakly curved gravity dual in the large-$N$ limit.
For completeness, we start with an explanation of the mABJM theory in terms of the fields
with manifest SU(2)$\times$SU(2)$\times$U(1) global symmetry, and
a summary of the supersymmetric vacua of the theory.

\subsection{mABJM theory with SU(2)$\times$SU(2)$\times$U(1) global symmetry}

The ${\cal N}=6$ mABJM theory is well established as a theory describing multiple
M2-branes~\cite{Hosomichi:2008jb,Gomis:2008vc} in the background of
a special kind of constant 4-form field strength and its dual seven-form field strength \cite{KKNT, Lambert:2009qw}.
There are several
methods for obtaining such a mass-deformed theory; for instance,  the ${\cal
N} = 1$ superfield formalism~\cite{Hosomichi:2008jb} and $D$-term and
$F$-term deformations in the ${\cal N} = 2$ superfield
formalism~\cite{Gomis:2008vc}. These different versions of
mass-deformed theories are actually equivalent since they are
connected by field redefinitions~\cite{Kim:2009ny}.
The action of this theory, which has U($N$)$\times$U($N$) gauge symmetry and SU(2)$\times$SU(2)$\times$U(1) global symmetry, is
\begin{align}\label{ZWdecom}
S=\int dx^3({\cal L}_0 + {\cal L}_{{\rm CS}}+{\cal L}_{{\rm ferm}}+{\cal L}_{\rm bos}),
\end{align}
where
\begin{align}\label{LLCS}
{\cal L}_{0}&={\rm tr}\Big(-D_\mu Z^\dagger_aD^\mu Z^a-D_\mu
W^{\dagger a}D^\mu W_a+i\xi^{\dagger}_a\gamma^\mu
D_\mu\xi^a+i\omega^{\dagger a}\gamma^\mu D_\mu\omega_a\Big),
\nonumber \\
{\cal L}_{{\rm CS}}&=\frac{k}{4\pi}\epsilon^{\mu\nu\rho} {\rm tr}
\Big(A_\mu\partial_\nu A_\rho+\frac{2i}3A_\mu A_\nu
A_\rho- \hat
A_\mu\partial_\nu \hat A_\rho -\frac{2i}3\hat A_\mu \hat A_\nu \hat
A_\rho\Big),
\nonumber\\
{\cal L}_{\rm ferm}&=-\frac{2\pi i}{k}{\rm tr}\Big[
(\xi^a\xi^\dagger_a-\omega^{\dagger
a}\omega_a)(Z^bZ^\dagger_b-W^{\dagger b}W_b) +2 (Z^a\xi^\dagger_a
+\omega^{\dagger a}W_a)(\xi^bZ^\dagger_b+W^{\dagger b}\omega_b)
\nonumber \\
&\hskip 2cm -(\xi^\dagger_a\xi^a-\omega_a\omega^{\dagger
a})(Z^{\dagger}_bZ^b-W_bW^{\dagger b})
-2(Z^\dagger_a\xi^a+\omega_aW^{\dagger a})(\xi^{\dagger}_bZ^b
+W_b\omega^{\dagger b})
\nonumber \\
&\hskip 2cm  + Z^a\omega_aZ^b\omega_b
+\xi^aW_a\xi^bW_b-2Z^aW_a\xi^b\omega_b -2Z^a\omega_a\xi^bW_b
\nonumber \\
&\hskip 2cm + \omega^{\dagger a}Z_a^\dagger\omega^{\dagger
b}Z_b^\dagger + W^{\dagger a}\xi_a^\dagger W^{\dagger
b}\xi_b^\dagger -2\omega^{\dagger a}\xi_a^\dagger W^{\dagger
b}Z_b^\dagger - 2 W^{\dagger a}\xi_a^\dagger\omega^{\dagger
b}Z_b^\dagger
\nonumber \\
&\hskip 2cm -\omega_aZ^a\omega_bZ^b-W_a\xi^aW_b\xi^b
+2\omega_aZ^aW_b\xi^b +2W_aZ^a\omega_b\xi^b
\nonumber \\
&\hskip 2cm  - Z_a^\dagger\omega^{\dagger
a}Z_b^\dagger\omega^{\dagger b} -\xi_a^\dagger W^{\dagger
a}\xi_b^\dagger W^{\dagger b} +2\xi_a^\dagger W^{\dagger a}
Z_b^\dagger\omega^{\dagger b} +2\xi_a^\dagger\omega^{\dagger a}
Z_b^\dagger W^{\dagger b}\Big]
\nn \\
&~~~~+i\mu\,{\rm tr}\big(\xi^\dagger_a\xi^a-\omega^{\dagger a}\omega_a\big),
\nonumber\\
{\cal L}_{\rm bos}&=-\left|\frac{2\pi}k\big(Z^bZ^\dagger_bZ^a-Z^aZ^\dagger_bZ^b+Z^aW_bW^{\dagger b}-W^{\dagger b}W_bZ^a\big)-\mu Z^a\right|^2\nn\\
&~~~-\left|\frac{2\pi}k\big(Z^\dagger_bZ^bW_a-W_aZ^bZ^\dagger_b+W_aW^{\dagger b}W_b-W_bW^{\dagger b}W_a\big)+\mu W_a\right|^2\nn\\
&~~~-\left|\frac{4\pi}k\big(W^{\dagger b}Z^\dagger_bW^{\dagger a}-W^{\dagger a}Z^\dagger_bW^{\dagger b}\big)\right|^2-\left|\frac{4\pi}k\big(Z^\dagger_aW^{\dagger b}Z^\dagger_b-Z^\dagger_bW^{\dagger b}_bZ^\dagger_a\big)\right|^2,
\end{align}
where $\left|{\cal O}\right|^2\equiv {\rm tr}\left({\cal O}^\dagger {\cal O}\right)$.
For later convenience, we have split the usual four transverse complex scalar and fermionic fields of the original SU(4)-invariant ABJM theory as
\begin{align}
Y^a=(Z^a,W^{\dagger a}),\quad \Psi_a =(\epsilon_{ab}\xi^b,-\epsilon_{ab}\omega^{\dagger b}),
\end{align}
where $a=1,2,3,4$ for the left-hand sides of these equations and $a,b=1,2$
for their right-hand sides.
The action is invariant under the ${\cal N}=6$ supersymmetry transformation,
\begin{align}\label{susyvar}
&\delta Z^a=i\bar\epsilon\xi^a+i\alpha^a_{~b}\omega^{\dagger b},\qquad \delta W^{\dagger a}=-i\epsilon\omega^{\dagger a}+i\beta^a_{~b}\xi^{b},
\nn\\
&\delta A_\mu=-\frac{2\pi}k\Big[\bar\epsilon\gamma_\mu\big(\xi^aZ^\dagger_a+W^{\dagger a}\omega_a\big)+\beta^b_{~a}\gamma^\mu\big(\xi^aW_b-Z^a\omega_b\big)+{\rm c.c.}\Big],
\nn\\
&\delta \hat A_\mu=-\frac{2\pi}k\Big[\bar\epsilon\gamma_\mu\big(Z^\dagger_a\xi^a+\omega_aW^{\dagger a}\big)+\beta ^b_{~a}\gamma^\mu\big(W_b\xi^a-\omega_bZ^a\big)+{\rm c.c.}\Big],
\nn\\
&\delta\xi^a=\epsilon\Big[\gamma^\mu D_\mu Z^a-\frac{4\pi}k\big(Z^{[a}Z^\dagger_bZ^{b]}+Z^{[a}W_bW^{\dagger b]}-2\epsilon^{ab}Z^{[1}Z^\dagger_bZ^{2]}\big)+\mu Z^a\Big]
\nn\\
&~~~~~~+\alpha^a_{~b}\Big[\gamma^\mu D_\mu W^{\dagger b}+\frac{4\pi}k\big(W^{\dagger [b}Z^\dagger_cZ^{c]}-W^{\dagger [b}W_cW^{\dagger c]}\big)+\mu W^{\dagger b}\Big]
\nn\\
&~~~~~~+\frac{8\pi}k\bar \epsilon\epsilon^{ab}W^{\dagger[ 1}Z^{\dagger}_bW^{\dagger 2]}
-\frac{8\pi}k\alpha^b_{~c}W^{\dagger [c}Z^\dagger_bZ^{a]},
\nn \\
&\delta\omega^{\dagger a}=\bar\epsilon\Big[-\gamma^\mu D_\mu W^{\dagger a}+\frac{4\pi}k\big(W^{\dagger[a}Z^\dagger_bZ^{b]}+W^{\dagger[a}W_bW^{\dagger b]}-2\epsilon^{ab}W^{\dagger[1}W_bW^{\dagger2]}\big)+\mu W^{\dagger a}\Big]
\nn\\
&~~~~~~~+\beta^a_{~b}\Big[\gamma^\mu D_\mu Z^{b}-\frac{4\pi}k\big(Z^{[b}Z^\dagger_cZ^{c]}-Z^{[b}W_cW^{\dagger c]}\big)-\mu Z^{b}\Big]
\nn\\
&~~~~~~~-\frac{8\pi}k\epsilon\epsilon^{ab}Z^{[1}W_bZ^{2]}-\frac{8\pi}k\beta^b_{~c}Z^{[c}W_bW^{\dagger a]},
\end{align}
where $A^{[a}B_bC^{c]}=\frac12(A^aB_bC^c-C^cB_bA^a)$ and
$\bar\epsilon=\epsilon^\ast$, $\beta^a_{~b}=(\alpha^b_{~a})^\ast$ are the supersymmetry parameters.
We also have $\epsilon^{ab}\beta^c_{~b}=-\epsilon^{bc}\alpha^a_{~b}$
from the reality condition of the supersymmetry parameters of the original
ABJM theory.

\subsection{Supersymmetric vacua of the mABJM theory}

In this subsection, we summarize the classical supersymmetric vacuum solutions of the mABJM theory, which were obtained in Ref. \cite{Gomis:2008vc}. The vacuum equations are obtained by setting the bosonic potential to zero, $V_{{\rm bos}} = - {\cal L}_{{\rm bos}}=0$, in Eq. \eqref{LLCS}.
Since $- {\cal L}_{{\rm bos}}$ is a sum of four absolute squares, we have four
equations that are cubic for $Z^{a}$ and $W_{a}$.
The first two are the $D$-term equations, while the last two are the $F$-term equations.
Suppose that the scalar fields are chosen to be
orthogonal to each other, $Z^aW_b=0$. Then the vacuum
equations are simplified by imposing this condition.
The $D$-term equations become
\begin{align}\label{Dtermeq}
Z^aZ^\dagger_bZ^b-Z^bZ^\dagger_bZ^a=-\frac{\mu k}{2\pi} Z^a,\qquad W^{\dagger a}W_b W^{\dagger b}
-W^{\dagger b} W_b W^{\dagger a}=\frac{\mu k}{2\pi} W^{\dagger a},
\end{align}
and the $F$-term equations are trivially satisfied as
\begin{align}\label{Ftermeq}
W_aZ^bW_b-W_bZ^bW_a=0,\qquad Z^bW_bZ^a-Z^aW_bZ^b=0.
\end{align}

The classical vacuum solutions of the $D$-term equations \eqref{Dtermeq}
have been found in the form of the GRVV matrices.
A convenient way of expressing each solution is to
represent it as a direct sum of two types of irreducible rectangular $n\times (n+1)$ matrices, ${\cal M}_a^{(n)}~(a=1,2)$, or a direct sum of the Hermitian conjugates of those rectangular matrices
$\bar{\cal M}_a^{(n)}$~\cite{Kim:2010mr,Cheon:2011gv}, which are
\begin{align}\label{mat-1}
{\cal M}_1^{(n)}&=\left(\begin{array}{cccccc}
\sqrt{n\!}\!\!\!&0&&&&\\&\!\sqrt{n\!-\!1} \!\!&\!0&&&\\
&&\ddots&\ddots&&\\&&&\sqrt{2}&0&\\&&&&1&0\end{array}\right),
\qquad
{\cal M}_2^{(n)}&=
\left(\begin{array}{cccccc}0&1&&&&\\&0&\sqrt{2}&&&\\ &&\ddots&\ddots&&\\
&&&0\!&\!\!\sqrt{n\!-\!1}\!&\\&&&&0&\!\!\!\sqrt{n\!}\end{array}
\right),
\end{align}
where $n=1,\cdots, N-1$.
In order to form $N\times N$ matrices, the direct sums of rectangular $n\times (n+1)$ matrices should also include certain numbers of empty rows and empty columns. Explicitly, the solutions are written
as follows\footnote{Similar vacuum solutions for the ${\cal N}=3$ mass-deformed CS matter theory
with generalized CS levels $k_1$ and $k_2$ were obtained in Ref. \cite{Kwon:2011nv}.}:
\begin{align}\label{ZW-vacua}
Z^a&=\sqrt{\frac{\mu k}{2\pi}}\left(\begin{array}{c}
\begin{array}{cccccc}\mathcal{M}_a^{(n_1)}\!\!&&&&&\\&\!\!\ddots\!&&&&\\
&&\!\!\mathcal{M}_a^{(n_i)}&&& \\ &&& {\bf 0}_{(n_{i+1}+1)\times n_{i+1}}
&&\\&&&&\ddots&\\&&&&&{\bf 0}_{(n_f+1)\times n_f}\end{array}\\
\end{array}\right),\nonumber
\end{align}
\begin{align}
W^{\dagger a}&=\sqrt{\frac{\mu k}{2\pi}}\left(\begin{array}{c}
\begin{array}{cccccc}{\bf 0}_{n_1\times (n_1+1)}&&&&&\\&\ddots&&&&\\
&&{\bf 0}_{n_i\times(n_i + 1)} &&&\\
&&& \bar{\mathcal M}_a^{(n_{i+1})}\!\!&&\\&&&&\!\!\ddots\!&\\
&&&&&\!\!\bar{\mathcal M}_a^{(n_f)}\end{array}\\
\end{array}\right).
\end{align}
The matrices ${\cal M}_a^{(n)}$ are called the $n$th block of the first type, while their Hermitian conjugates $\bar{\cal M}_a^{(n)}$ are called the $n$th block of the second type.
Since the nonzero components of $Z^a$ and $W^{\dagger a}$ belong to different blocks, the product $Z^aW_b$ always vanishes and the $F$-term vacuum equations
 \eqref{Ftermeq} are automatically satisfied. The solutions in Eq. \eqref{ZW-vacua} also satisfy the $D$-term equations \eqref{Dtermeq} because the rectangular blocks ${\cal M}_a^{(n)}$ solve the equations block by block as follows:
\begin{align}
\sum_{b=1}^2\big({\cal M}_a^{(n)}\bar{\cal M}_b^{(n)}{\cal M}_b^{(n)}-{\cal M}_b^{(n)}\bar{\cal M}_b^{(n)}{\cal M}_a^{(n)}\big)=-{\cal M}_a^{(n)}.
\end{align}

One occupation number, $N_n$ with $n=0,1,\cdots, N-1$, denotes the number of blocks of ${\cal M}_a^{(n)}$ contained in $Z^a$, and another occupation number, $N_n'$, denotes the number of blocks of $\bar{\cal M}_a^{(n)}$ contained in $W^{\dagger a}$. In this notation $N_0$ is the number of empty columns while $N_0'$ is the number of empty rows. Since $Z^a$ and $W^{\dagger a}$ are $N\times N$ matrices, the occupation numbers should satisfy the following constraints:
\begin{align}\label{levelmatch}
\sum_{n=0}^{N-1}\big[nN_n+(n+1)N_n'\big]=N,\qquad
\sum_{n=0}^{N-1}\big[(n+1)N_n+nN_n'\big]=N.
\end{align}
At the quantum level, only a subset of these classical vacuum solutions remains supersymmetric, and the occupation numbers for the quantum-level supersymmetric vacua are further constrained by the value of
the CS level $k$ as
\begin{align}\label{susyvacuum}
0\le N_n\le k,\qquad 0\le N_n'\le k
\end{align}
for every $n$~\cite{Kim:2010mr}.

\section{Abelian projections}

\subsection{${\cal N}=4$ Abelian projection}\label{N=4abelian}
The Abelian projection of the mABJM theory is implemented by making an ansatz for the dynamical fields. They are given by two $N\times N$ matrices, $S^1$ and $S^2$, and are expressed by direct sums of the rectangular blocks, ${\cal M}_1^{(n)}$ and ${\cal M}_2^{(n)}$, respectively. Since all ${\cal M}_a^{(n)}$'s are $n\times (n+1)$ matrices, their direct sum always forms a rectangular matrix with fewer rows than  columns. Therefore, in order to form the $N\times N$ matrices, the direct sums should contain a certain number of empty rows. In a general ansatz a certain number of empty columns is also allowed. More precisely, if $N_n$ is the number of ${\cal M}_a^{(n)}$ blocks contained in $S^a$, $N_0'$ empty rows should be included in order to form the $N\times N$ matrices. It is also possible to include $N_0$ empty columns. Then the occupation numbers must satisfy the following constraints:
\begin{align}\label{vacNN}
\sum_{n=1}^{N-1}nN_n+N_0'=N,\qquad \sum_{n=1}^{N-1}(n+1)N_n+N_0=N.
\end{align}
The $S^a$ matrices of \(a=1\) or \(2\) constructed in this manner satisfy the following properties:
\begin{align}\label{Saprop}
S^a S_b^\dagger S^b - S^b S_b^\dagger S^a = -S^a,\quad
{\rm tr}(S^1S_{1}^\dagger) = {\rm tr}(S^2S_{2}^\dagger)=\sum_{n=0}^{N-1}N_n\frac{n(n+1)}2.
\end{align}

In order to obtain the Abelian projected theory of the mABJM theory,
one can consider the following ansatz~\cite{Mohammed:2012gi}:
\begin{align}\label{abeans}
&Z^a = \phi_a(x) S^a, \qquad W^{\dagger a} = \rho_a(x)S^a
\nn \\
& \xi^a = \psi_a(x) S^a, \qquad \omega^{\dagger a} =\chi_a(x)S^a,
\nn \\
& A^\mu = a^\mu_2(x) S^1S_1^\dagger+a^\mu_1(x)S^2S_2^\dagger,\nn\\
&\hat A^\mu = a_2^\mu(x) S_1^\dagger S^1+a_1^\mu(x) S_2^\dagger S^2,
\end{align}
where here and from now on we understand that there is no summation on the repeated indices ($a,b$), unless stated otherwise. We notice that since for this ansatz the product $Z^aW_b$ is nonvanishing---even when the coefficient fields $\phi_a(x)$ and $\rho_a(x)$ are made constant---Eq. \eqref{abeans} does not satisfy the vacuum equations of the mABJM theory.
Therefore, this ansatz cannot be considered a fluctuation on a supersymmetric vacuum. By inserting the ansatz \eqref{abeans} into the mass-deformed Lagrangian in Eq. \eqref{ZWdecom}, and by using the properties of the $S^a$ matrices in Eq. \eqref{Saprop}, we obtain
\begin{align}\label{N=4CS}
{\cal L}_{\rm tot} = \beta\bigg\{&-({\cal D}_\mu\phi_a)^\dagger {\cal D}^\mu \phi_a-({\cal D}_\mu\rho_a)^\dagger {\cal D}^\mu \rho_a + i\bar\psi_a\gamma^\mu {\cal D}_\mu\psi_a+ i\bar\chi_a\gamma^\mu {\cal D}_\mu\chi_a\nn\\
&+\frac{k}{4\pi}\epsilon_{\mu\nu\rho} \big(a_1^\mu\partial^\nu a_2^\rho+a_2^\mu\partial^\nu a_1^\rho\big)\nn\\
& -\frac{2\pi i}{k}\Big[(|\psi_1|^2-|\chi_1|^2)(|\phi_2|^2-|\rho_2|^2)+(|\psi_2|^2-|\chi_2|^2)(|\phi_1|^2-|\rho_1|^2)\nn\\
&~~~~~~~~~~+2(\phi_1\bar\psi_1+\chi_1\bar\rho_1)(\psi_2\bar\phi_2+\rho_2\bar\chi_2)
+2(\phi_2\bar\psi_2+\chi_2\bar\rho_2)(\psi_1\bar\phi_1+\rho_1\bar\chi_1)\Big]
\nn \\
&-\frac{4\pi i}k\Big[\phi_1\phi_2\bar\chi_1\bar\chi_2+\bar\phi_1\bar\phi_2\chi_1\chi_2
+\bar\rho_1\bar\rho_2\psi_1\psi_2+\rho_1\rho_2\bar\psi_1\bar\psi_2-\phi_1\bar\rho_1\psi_2\bar\chi_2-\bar\phi_1\rho_1\bar\psi_2\chi_2\nn\\
&~~~~~~~-\phi_1\bar\rho_2\bar\chi_1\psi_2-\bar\phi_1\rho_2\chi_1\bar\psi_2-\phi_2\bar\rho_2\psi_1\bar\chi_1-\bar\phi_2\rho_2\bar\psi_1\chi_1
-\phi_2\bar\rho_1\psi_1\bar\chi_2-\bar\phi_2\rho_1\bar\psi_1\chi_2\Big]\nn\\
& -\frac{4\pi^2}{k^2}\Big[(|\phi_1|^2+|\rho_1|^2)(|\phi_2|^2-|\rho_2|^2)^2
+(|\phi_2|^2+|\rho_2|^2)(|\phi_1|^2-|\rho_1|^2)^2\Big]\nn\\
&-\frac{16\pi^2}{k^2}\Big[(|\phi_1|^2+|\phi_2|^2)|\rho_1|^2|\rho_2|^2+|\phi_1|^2|\phi_2|^2(|\rho_1|^2+|\rho_2|^2)\Big]\nn\\
&+i\mu\big(\bar\psi_a\psi_a-\bar\chi_a\chi_a\big)-\mu^2\big(\bar\phi_a\phi_a+\bar\rho_a\rho_a\big)+\frac{8\pi\mu}k\big(|\phi_1|^2|\phi_2|^2
-|\rho_1|^2|\rho_2|^2\big)\bigg\},
\end{align}
where a summation over repeated indices is implied in this Abelian Lagrangian, and
\begin{align}
\beta=\sum_{n=0}^{N-1}N_n\frac{n(n+1)}2,\quad \bar X_a\equiv X_a^*,\quad {\cal D}^\mu X_a \equiv \partial^\mu X_a+ia^\mu_aX_a.
\end{align}
The Abelian projection considered in Ref.~\cite{Mohammed:2012gi} corresponds to a special case of the ansatz \eqref{abeans} in which $N_{N-1}=1$ and $N'_0=1$ are the only vanishing occupation numbers
and all other occupation numbers are zero. In this case the overall constant in the Lagrangian in Eq. \eqref{N=4CS} becomes $\beta=N(N-1)/2$.

The Abelian projection is also applied to the supersymmetry transformation rules in Eq. \eqref{susyvar}. It turns out that the Abelian theory obtained in Eq. \eqref{N=4CS} preserves only ${\cal N}=4$ supersymmetry. We begin with the Abelian projection of the supersymmetry variation of one scalar field,
\begin{align}
\delta Z^1=i\bar\epsilon\xi^1+i\alpha^1_{~1}\omega^{\dagger 1}+i\alpha^1_{~2}\omega^{\dagger 2}
\,\,\longrightarrow\,\,
\delta\phi_1S^1=i\bar\epsilon\psi_1S^1+i\alpha^1_{~1}\chi_1S^1+i\alpha^1_{~2}\chi^2S^2.
\end{align}
Recalling that the $S^1$ and $S^2$ are linearly independent matrices, the above equation is satisfied only if we set $\alpha^1_{~2}$ to zero. Now we set $\alpha^1_{~1}\equiv-\eta$ and have
\begin{align}
\delta\phi_1=i\bar\epsilon\psi_1-i\eta\chi_1.
\end{align}
Similarly, the application of the Abelian projection to the supersymmetry variations of the remaining bosonic and fermionic fields in Eq. \eqref{susyvar} results in the following ${\cal N}=4$ supersymmetry transformation rules:
\begin{align}\label{absusy}
&\delta\phi_1=i\bar\epsilon\psi_1-i\eta\chi_1,\qquad \delta\rho_1=-i\epsilon\chi_1-i\bar\eta\psi_1,\nn\\
&\delta\phi_2=i\bar\epsilon\psi_2+i\bar\eta\chi_2,\qquad \delta\rho_2=-i\epsilon\chi_2+i\eta\psi_2,\nn\\
&\delta a_1^\mu=-\frac{2\pi}k\Big[\bar\epsilon\gamma^\mu(\psi_2\bar\phi_2+\bar\chi_2\rho_2)+\eta\gamma^\mu(\psi_2\bar\rho_2-\bar\chi_2\phi_2)+{\rm c.c.}\Big],\nn\\
&\delta a_2^\mu=-\frac{2\pi}k\Big[\bar\epsilon\gamma^\mu(\psi_1\bar\phi_1+\bar\chi_1\rho_1)-\bar\eta\gamma^\mu(\psi_1\bar\rho_1-\bar\chi_1\phi_1)+{\rm c.c.}\Big],\nn\\
&\delta\psi_1=\epsilon\Big[\gamma^\mu {\cal D}_\mu\phi_1-\frac{2\pi}k\phi_1(|\phi_2|^2-|\rho_2|^2)+\mu\phi_1\Big]-\frac{4\pi}k\bar\epsilon\bar\phi_2\rho_1\rho_2\nn\\
&~~~~~~~+\eta\Big[\gamma^\mu {\cal D}_\mu\rho_1+\frac{2\pi}k\rho_1(|\phi_2|^2-|\rho_2|^2)-\mu\rho_1\Big]-\frac{4\pi}k\bar\eta\phi_1\bar\phi_2\rho_2,\nn\\
&\delta\psi_2=\epsilon\Big[\gamma^\mu {\cal D}_\mu\phi_2-\frac{2\pi}k\phi_2(|\phi_1|^2-|\rho_1|^2)+\mu\phi_2\Big]-\frac{4\pi}k\bar\epsilon\bar\phi_1\rho_1\rho_2\nn\\
&~~~~~~~+\bar\eta\Big[\gamma^\mu {\cal D}_\mu\rho_2-\frac{2\pi}k\rho_2(|\phi_1|^2-|\rho_1|^2)+\mu\rho_2\Big]+\frac{4\pi}k\eta\bar\phi_1\phi_2\rho_1,\nn\\
&\delta\chi_1=-\bar\epsilon\Big[\gamma^\mu {\cal D}_\mu\rho_1+\frac{2\pi}k\rho_1(|\phi_2|^2-|\rho_2|^2)-\mu\rho_1\Big]+\frac{4\pi}k\epsilon\phi_1\phi_2\bar\rho_2\nn\\
&~~~~~~~-\bar\eta\Big[\gamma^\mu {\cal D}_\mu\phi_1+\frac{2\pi}k\phi_1(|\phi_2|^2-|\rho_2|^2)-\mu\phi_1\Big]-\frac{4\pi}k\eta\phi_2\rho_1\bar\rho_2,\nn\\
&\delta\chi_2=-\bar\epsilon\Big[\gamma^\mu {\cal D}_\mu\rho_2+\frac{2\pi}k\rho_2(|\phi_1|^2-|\rho_1|^2)-\mu\rho_2\Big]+\frac{4\pi}k\epsilon\phi_1\phi_2\bar\rho_1\nn\\
&~~~~~~~+\eta\Big[\gamma^\mu {\cal D}_\mu\phi_2+\frac{2\pi}k\phi_2(|\phi_1|^2-|\rho_1|^2)-\mu\phi_2\Big]+\frac{4\pi}k\bar\eta\phi_1\bar\rho_1\rho_2.
\end{align}
With a straightforward but tedious calculation, the Abelian action in Eq. \eqref{N=4CS} is shown to be invariant under this ${\cal N}=4$ supersymmetry transformation.

The supersymmetry enhancement for $k=1,2$ in the (m)ABJM theory is realized by using the
monopole operators~\cite{Gustavsson:2009pm}, which relate the bifundamental fields and the antibifundamental
fields. The GRVV matrices $S^a$ and $S_a^\dagger$ in the ansatz \eqref{abeans},
which represent the bifundamental and antibifundamental representations, respectively,
are independent of each other.  For this reason, the monopole operators related to the supersymmetry
enhancement do not exist in Abelian projected theories. Therefore, no supersymmetry enhancement is expected in these Abelian theories. In general, however, monopole operators~\cite{'tHooft:1977hy} in three-dimensional gauge
theories create a U(1) magnetic flux in the gauge group under consideration.
The Abelian projection can be considered as a special way to extract the U(1) sector
in which monopole operators can live. Therefore, monopole operators that
are not related to the supersymmetry enhancement can exist in the Abelian projected theories.

The next point to clarify is that the Abelian projection is a consistent truncation of the original mABJM theory. In Ref. \cite{Mohammed:2012gi} this was argued in the sense that the solutions to the equations of motion of the truncated theory are also solutions to the original theory. This is indeed the case, because by using the properties of the $S^a$ matrices in Eq. \eqref{Saprop} one can easily verify that $D_\mu D^\mu(X_aS^a)=({\cal D}_\mu{\cal D}^\mu X_a)S^a$. Using the same set of properties, other expressions in the equations of motion of the mABJM theory involving products of more than two $S^a$ matrices can be reduced to expressions involving only one matrix. As a result of these simplifications the equations of motion of the original mABJM theory can be reduced to the equations of motion of the Abelian projected theory using the ansatz \eqref{abeans}.

 As a matter of fact, the above criteria of a consistent truncation allows a further truncation of the Abelian theory to a simpler Abelian theory involving fewer scalar fields. In particular, a truncation involving
 two scalar fields can be obtained by setting any two of the four scalar fields and the corresponding fermionic fields to zero. There are three inequivalent such truncations:
(i) $\phi_2=\rho_2=0$,
(ii) $\phi_1=\rho_2=0$, and
(iii) $\rho_1=\rho_2=0$.
Other cases are related to these three cases by field renaming~\cite{Mohammed:2012gi}.

In case (i) the truncated action is
\begin{align}\label{N=4CS2}
{\cal L}_{\rm tot} = \beta\Big[&-({\cal D}_\mu\phi_1)^\dagger {\cal D}^\mu \phi_1-({\cal D}_\mu\rho_1)^\dagger {\cal D}^\mu \rho_1 + i\bar\psi_1\gamma^\mu {\cal D}_\mu\psi_1+ i\bar\chi_1\gamma^\mu {\cal D}_\mu\chi_1\nn\\
&+\frac{k}{4\pi}\epsilon_{\mu\nu\rho} a_2^\mu f_1^{\nu\rho}+i\mu\big(|\psi_1|^2-|\chi_1|^2\big)-\mu^2\big(|\phi_1|^2+|\rho_1|^2\big)\Big],
\end{align}
where $f_i^{\mu\nu}=\partial^\mu a_i^\nu-\partial^\nu a_i^\mu$ are the field strengths of the Abelian gauge fields. The gauge field $a_2^\mu$ is an auxiliary field and its equation of motion is $f_1^{\mu\nu}=0$, which means $a_1^{\mu}$ is a pure gauge degree of freedom. Therefore, the truncation results in a theory of two massive bosonic and two massive fermionic fields without self-interactions of the scalar fields and Yukawa-type interactions. This theory is invariant under the ${\cal N}=4$ supersymmetry transformation.

In case (ii) the truncated action is
\begin{align}\label{case2}
{\cal L}= \beta\bigg[& ({\cal D}_\mu\phi_2)^\dagger {\cal D}^\mu \phi_2-({\cal D}_\mu\rho_1)^\dagger {\cal D}^\mu \rho_1 + i\bar\psi_2\gamma^\mu {\cal D}_\mu\psi_2+ i\bar\chi_1\gamma^\mu {\cal D}_\mu\chi_1\nn\\
&+\frac{k}{4\pi}\epsilon_{\mu\nu\rho} \big(a_1^\mu\partial^\nu a_2^\rho+a_2^\mu\partial^\nu a_1^\rho\big)+ \frac{4\pi^2 }{k^2}\Big(|\phi_2|^2|\rho_1|^2 + \frac{\mu^2 k^2}{4\pi^2}\Big)
\left(|\phi_2|^2 + |\rho_1|^2\right)\\
&+\frac{2\pi i}{k}\left( |\phi_2|^2|\chi_1|^2 +
|\rho_1|^2 |\psi_2|^2 - 2 \chi_1\bar\rho_1\psi_2\bar\phi_2
-2 \phi_2\bar\psi_2\rho_1\bar\chi_2 \right)
+ i\mu \left(\bar\psi_2\psi_2 - \bar\chi_1\chi_1\right)\bigg]\nn.
\end{align}
Only ${\cal N}=2$ supersymmetry
remains, with the supersymmetry transformation rules
\begin{align}
&\delta\psi_2 = i\bar\epsilon\psi_2,\quad \delta\rho_1 = -i \epsilon\chi_1,
\nn \\
&\delta a_1^\mu = -\frac{2\pi}{k} \left(\bar\epsilon \gamma^\mu\psi_2\bar\phi_2
+  \bar\psi_2\gamma^\mu\epsilon\phi_2\right),\quad\delta a_2^\mu = -\frac{2\pi}{k} \left(\bar\epsilon \gamma^\mu\bar\chi_1\rho_1
+ \chi_1\gamma^\mu\epsilon\bar\rho_1\right),
\nn \\
&\delta\psi_2 = \epsilon\left(\gamma^\mu  {\cal D}_\mu\phi_2 + \frac{2\pi}{k}\phi_2|\rho_1|^2
+ \mu \phi_2\right),\quad\delta\chi_1 = -\bar\epsilon\left(\gamma^\mu  {\cal D}_\mu\rho_1 + \frac{2\pi}{k}\rho_1|\phi_2|^2
- \mu \rho_1\right).
\end{align}
The bosonic potential in Eq. \eqref{case2} has no Higgs vacuum, and as such nontopological vortex-type BPS solutions are supported \cite{Mohammed:2012gi}.

Case (iii) is the most interesting because of its relevance in describing some
condensed matter models~\cite{Mohammed:2012gi}. The truncated action is
\begin{align}\label{case3}
{\cal L} = \beta\bigg[&-({\cal D}_\mu\phi_a)^\dagger {\cal D}^\mu \phi_a+ i\bar\psi_a\gamma^\mu {\cal D}_\mu\psi_a+\frac{k}{4\pi}\epsilon_{\mu\nu\rho} \big(a_1^\mu\partial^\nu a_2^\rho+a_2^\mu\partial^\nu a_1^\rho\big)\nn\\
& -\frac{2\pi i}{k}\Big(|\psi_1|^2|\phi_2|^2+|\psi_2|^2|\phi_1|^2+2\phi_1\bar\psi_1\psi_2\bar\phi_2
+2\phi_2\bar\psi_2\psi_1\bar\phi_1\Big)
\nn \\
& -\frac{4\pi^2}{k^2}\Big(|\phi_1|^2|\phi_2|^4
+|\phi_2|^2|\phi_1|^4\Big)+i\mu\bar\psi_a\psi_a-\mu^2\bar\phi_a\phi_a+\frac{8\pi\mu}k|\phi_1|^2|\phi_2|^2
\bigg].
\end{align}
This action is also invariant under the ${\cal N}=2$ supersymmetry transformation using the transformation rules obtained from Eq. \eqref{absusy} by setting the supersymmetry parameter $\eta$ to zero.

\subsection{Comments on the gravity dual of the ${\cal N}=4$ Abelian theories}\label{difficulty}
Ans\"atze of Abelian projections similar to Eq. \eqref{abeans} were commonly used in the literature to implement the Abelian projection of the ABJM theory~\cite{Arai:2008kv,Kawai:2009rc,Mohammed:2012gi}. However, as it was pointed out in the pervious subsection, these ans\"atze do not satisfy the vacuum equations of the mABJM theory unless we set $\phi_a$ or $\rho_a$ to zero. If the latter is not the case, the ans\"atze can not be considered as fluctuations on the supersymmetric vacua of the mABJM theory. Then the identification of the gravity dual is unclear, since there is no known geometry which is dual to the configurations given
by Eq. \eqref{abeans}.
Due to this reason, for the ${\cal N}=4$ Abelian gauge theories, we cannot use the holographic duality relations between the supersymmetric vacua of the mABJM theory and the LLM geometries.
One has to find some other ways to identify the gravity dual of the ${\cal N}=4$ Abelian gauge theories.

When $\phi_a$ or $\rho_a$ are set to zero, one can identify the dual gravity solutions: however, there are still some obstacles involved with these solutions. We discuss these obstacles in this subsection, and we construct an ${\cal N}=2$ Abelian projection based on the vacua of the mABJM theory in the next subsection.

Let us consider a particular vacuum solution with occupation numbers
$\{N_n,N_n'\}$ satisfying the supersymmetric condition \eqref{susyvacuum}.
In Ref. \cite{Cheon:2011gv}, the $\mathbb Z_k$ quotient of the LLM geometry~\cite{Lin:2004nb} is identified as the
holographic dual of the vacua of the mABJM theory.
This geometry is represented in terms of an infinite strip of white and black regions in Fig.~1, which are called droplets. In this representation the two colors correspond to $\pm\frac12$ boundary values of the function that characterizes the metric; however, we leave its clarification until Sec. 4.2. The strip is divided into excitation levels of length $k$, which are labeled by non-negative integers $n=0,1,2,\cdots$ starting at the Fermi level ($E_{\rm F}$). In this strip representation, the occupation number $N_n$ corresponds to the length of the black region in the $n$th level above the Fermi level, while $N_n'$ corresponds to that of the white region in the $n$th level below the Fermi level. Since the length of the black/white strip in a given level cannot be bigger than the length of the level itself, the occupation numbers should satisfy the condition \eqref{susyvacuum}, which is also required in order to have supersymmetric vacua \cite{Kim:2010mr}.

An alternative representation of those gravity solutions is expressed in terms of a Young diagram. For the LLM geometry corresponding to the $k=1$ case, the lengths of the white/black strips are equal to those of the horizontal/vertical edges of the Young diagram, as shown in Figs.~1(b) and (c).
The Young diagrams for small-curvature solutions should all have very long edges and a very large total number of boxes~\cite{Lin:2004nb}. In other words, a small-curvature geometry is represented by a Young diagram whose shape is a square or almost a square. If the Young diagram has a large number of corners, it is most likely that some of the edges are short and the geometry includes highly curved regions. In Sec. 4 we will discuss the structure of LLM geometry and its various limits.

Based on the discussion in the above two paragraphs, we notice that there are two obstacles with the dual gravity solutions to the Abelian projection ansatz in Eq. \eqref{abeans} with $\phi_a$ or $\rho_a$ set to zero. The first one is the fact that since the ansatz is based on matrices that are built from only $n\times (n+1)$ rectangular blocks, we have to add $N_0'$ empty rows, where $N_0'$ is equal to the total number of rectangular blocks used. Unless $n$ is of the order of $N$, one has to use a large number of blocks to form an $N\times N$ matrix, and as a result $N_0'$ can be larger than $k$, in contrast to the requirement from supersymmetric vacuum solutions \eqref{susyvacuum}. The second obstacle is related to the cases when the rectangular building blocks have sizes of the order of $N$, so that only higher-level occupation numbers (such as $N_{N-1},N'_{N-1},\cdots$) are nonzero. Since the $N\times N$ matrices representing the vacuum solutions can not contain more than one ${\cal M}_a^{(N-1)}$ or $\bar{\cal M}_a^{(N-1)}, \cdots$, these nonvanishing occupation numbers are also of the order of one unit. Therefore, in the droplet representation, the length of either the black strip or the white strip is of the order of one unit.
In these cases, the corresponding Young diagrams will have some edges that have lengths of the order $N$ and others with lengths of the order of one unit. For instance, in the case of Ref. \cite{Mohammed:2012gi}, the Young diagram is a $1\times N$ rectangle. As pointed out before, for a Young diagram with short edges, the dual geometry includes highly curved regions and the gravity approximation of the fluctuations on such a geometry cannot be trusted.
However, one can still consider the identification of the solitonic
objects in the dual gravity for the Young diagram with a $1\times N$
rectangle~\cite{Mohammed:2012gi,Auzzi:2009es}.

\begin{figure}
\centerline{\epsfig{figure=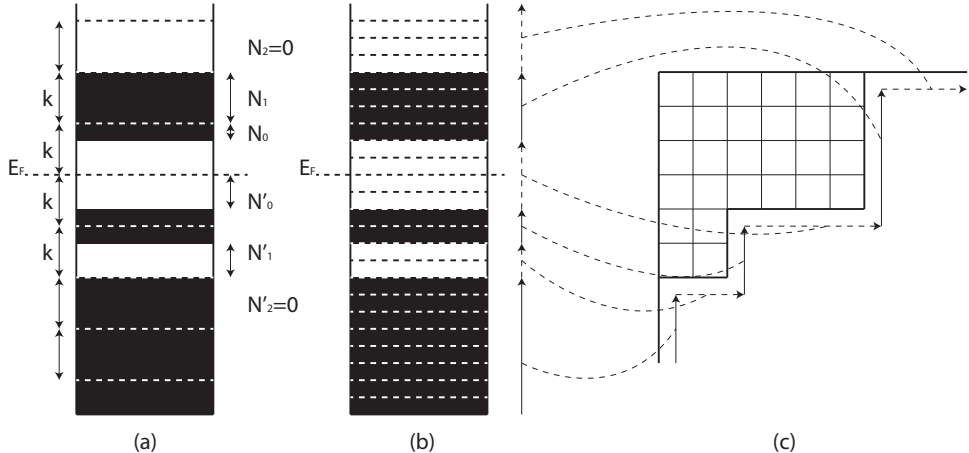,height=60mm}}
\caption{
\small (a) A droplet representation of the $\mathbb{Z}_k$ quotient of the LLM geometry.
The horizontal lines do not correspond to any coordinate, but rather are fictitious lines added for clarity. (b) A droplet representation for the $k=1$ case. (c) The Young diagram corresponding to the droplet (b).
}
\label{Fig1}
\end{figure}
The prime motivation for considering the Abelian truncation of the mABJM theory in Ref. \cite{Mohammed:2012gi}
is to achieve a better understanding of the gauge/gravity duality in condensed
matter systems.
The authors obtained the Abelian Higgs theory---the relativistic version of Landau-Ginzburg theory---through the Abelian projection of the mABJM theory. It was suggested that the strong-coupling regime of this theory can be investigated by using the gauge/gravity duality, though this point was not explicitly addressed. However, due to the two obstacles discussed in the pervious paragraph, it seems that the Abelian projection ansatz \eqref{abeans} is not suitable for the consideration of a weakly curved gravity dual\footnote{An alternative way of  realizing the gauge/gravity duality in condensed matter systems was suggested by adding fundamental fields to the ABJM theory~\cite{Murugan:2013jm}.}.
Since the weakly curved limit is the only region where one can trust the gravity approximation of the dual theory, it is not clear if it is practically possible to implement the idea of the gauge/gravity duality for condensed matter
models obtained from the ${\cal N}=4$ Abelian theory of Sec. \ref{N=4abelian}. In the next subsection, we will introduce an alternative truncation ansatz to circumvent these obstacles.

\subsection{${\cal N}=2$ Abelian projection}
In this subsection, we consider an Abelian projection of the mABJM theory where the ansatz for the dynamical fields is made in terms of four matrices, $S^a$ and $T^a$. The ansatz is identified by the occupation numbers $\{N_n,N_n'\}$, where $N_n$ is the number of blocks ${\cal M}_a^{(n)}$ contained in $S^a$, and $N_n'$ is the number of blocks $\bar {\cal M}_a^{(n)}$ contained in $T^a$. The blocks ${\cal M}_a^{(n)}$ and $\bar {\cal M}_a^{(n)}$ must be fitted as in Eq. \eqref{ZW-vacua} so that the products of $S^a$ with $T_b^\dagger$ are zero and the occupation numbers satisfy Eq. \eqref{levelmatch}.
It is easy to see that
\begin{align}\label{STaprop}
&S^a S_b^\dagger S^b - S^b S_b^\dagger S^a = -S^a,\qquad T^a T_b^\dagger T^b - T^b T_b^\dagger T^a = T^a, \nn\\
&{\rm tr}(S^aS_{a}^\dagger)= \sum_{n=0}^{N-1}N_nn(n+1),\qquad {\rm tr}(T^aT_{a}^\dagger)= \sum_{n=0}^{N-1}N_n'n(n+1).
\end{align}
The occupation numbers should also satisfy the supersymmetric vacuum condition \eqref{susyvacuum}.

We introduce the following truncation ansatz:
\begin{align}\label{abeans2}
&Z^a = \phi_a(x) S^a, \qquad W^{\dagger a} = \rho_a(x)T^a,
\nn \\
& \xi^a = \psi_a(x) S^a, \qquad \omega^{\dagger a} =\chi_a(x)T^a,
\nn \\
& A^\mu = a^\mu_2(x) S^1S_1^\dagger+a^\mu_1(x)S^2S_2^\dagger+b^\mu_2(x) T^1T_1^\dagger+b^\mu_1(x)T^2T_2^\dagger,\nn\\
&\hat A^\mu = a_2^\mu(x) S_1^\dagger S^1+a_1^\mu(x) S_2^\dagger S^2+b_2^\mu(x) T_1^\dagger T^1+b_1^\mu(x) T_2^\dagger T^2,
\end{align}
where $b_i^\mu$ are a new pair of Abelian gauge fields. Since all the vacuum equations are satisfied by this ansatz when $\phi_a$ and $\rho_a$ are set to $\sqrt{\frac{\mu k}{2\pi}}$, this ansatz can be considered as a special fluctuation on the supersymmetric vacua of the mABJM theory when we set
\begin{align}
\phi_a(x)=\sqrt{\frac{\mu k}{2\pi}}+\tilde\phi_a(x),\quad \rho_a(x)=\sqrt{\frac{\mu k}{2\pi}}+\tilde\rho_a(x).
\end{align}
Inserting the ansatz \eqref{abeans2} into the mass-deformed Lagrangian in Eq. \eqref{ZWdecom}, and using the properties of the $(S^a,T^a)$ matrices given in Eq.~\eqref{STaprop}, we obtain
\begin{align}\label{N=2CS-2}
{\cal L}_{\rm tot} = &\beta\bigg\{-({\cal D}_\mu\phi_a)^\dagger {\cal D}^\mu \phi_a + i\bar\psi_a\gamma^\mu {\cal D}_\mu\psi_a
+\frac{k}{4\pi}\epsilon_{\mu\nu\rho} \big(a_1^\mu\partial^\nu a_2^\rho+a_2^\mu\partial^\nu a_1^\rho\big)
\nn\\
&~~~~~~ -\frac{2\pi i}{k}\Big[|\psi_1|^2|\phi_2|^2+|\psi_2|^2|\phi_1|^2+2\big(\phi_1\bar\phi_2\bar\psi_1\psi_2
+\phi_2\bar\phi_1\bar\psi_2\psi_1\big)\Big]
\nn\\
&~~~~~~ -\frac{4\pi^2}{k^2}\big(|\phi_1|^4|\phi_2|^2+|\phi_1|^2|\phi_2|^4\big)
+i\mu\bar\psi_a\psi_a-\mu^2\bar\phi_a\phi_a
+\frac{8\pi\mu}k |\phi_1|^2|\phi_2|^2\bigg\}
\nn \\
& +\beta'\bigg\{-\big({\cal D}_\mu\rho_a)^\dagger {\cal D}^\mu \rho_a
+ i\bar\chi_a\gamma^\mu {\cal D}_\mu\chi_a
-\frac{k}{4\pi}\epsilon_{\mu\nu\rho} \big(b_1^\mu\partial^\nu b_2^\rho+b_2^\mu\partial^\nu b_1^\rho\big)
\nn\\
&~~~~~~+\frac{2\pi i}{k}\Big[|\chi_1|^2|\rho_2|^2+|\chi_2|^2|\rho_1|^2+2\big(\bar\rho_1\rho_2\chi_1\bar\chi_2
+\bar\rho_2\rho_1\chi_2\bar\chi_1\big)\Big]
\nn \\
&~~~~~~ -\frac{4\pi^2}{k^2}
\big(|\rho_1|^4|\rho_2|^2+|\rho_1|^2|\rho_2|^4\big)
-i\mu\bar\chi_a\chi_a-\mu^2\bar\rho_a\rho_a
-\frac{8\pi\mu}k |\rho_1|^2|\rho_2|^2\bigg\},
\end{align}
where
\begin{align}
&\beta=\sum_{n=0}^{N-1}N_n\frac{n(n+1)}2,\quad \beta'=\sum_{n=0}^{N-1}N_n'\frac{n(n+1)}2,
\nn \\
&{\cal D}^\mu (\phi_a,\psi_a) \equiv (\partial^\mu +ia^\mu_a)(\phi_a,\psi_a),
\qquad {\cal D}^\mu (\rho_a,\chi_a) \equiv (\partial^\mu -ib^\mu_a)(\rho_a,\chi_a).
\end{align}

As they appear here, Abelian projected theories obtained with different choices of the occupation numbers are the same except for the difference in the overall factor $\beta$ or $\beta'$. On the other hand, as was discussed in Refs. \cite{Kim:2010mr,Cheon:2011gv}, different supersymmetric vacua obtained by different choices of the occupation numbers are related to different dual gravity backgrounds. This fact suggests that the overall factors in the Abelian projected theories have a more important physical meaning than it at first appears. Below, we will consider a special choice of the occupation numbers and discuss the physical meanings of this choice in relation to the background geometry of the gravity dual.

The application of the Abelian projection ansatz \eqref{abeans2} to the supersymmetry transformation rules in Eq. \eqref{susyvar} preserves only ${\cal N}=2$ supersymmetry. The resulting supersymmetry transformation rules are
\begin{align}\label{absusy2}
&\delta\phi_1=i\bar\epsilon\psi_1,\quad \delta\phi_2=i\bar\epsilon\psi_2,\quad \delta\rho_1=-i\epsilon\chi_1,\quad \delta\rho_2=-i\epsilon\chi_2,\nn\\
&\delta a_1^\mu=-\frac{2\pi}k
\left(\bar\epsilon\gamma^\mu\psi_2\bar\phi_2
+\bar\psi_2\gamma^\mu\epsilon\phi_2\right),\qquad \delta a_2^\mu=
-\frac{2\pi}k\left(\bar\epsilon\gamma^\mu\psi_1\bar\phi_1
+\bar\psi_1\gamma^\mu\epsilon\phi_1\right),
\nn\\
&\delta b_1^\mu=-\frac{2\pi}k\left(\bar\epsilon\gamma^\mu\bar\chi_2\rho_2
+\chi_2\gamma^\mu\epsilon\bar\rho_2\right), \qquad
\delta b_2^\mu=-\frac{2\pi}k\left(\bar\epsilon\gamma^\mu\bar\chi_1\rho_1
+\chi_1\gamma^\mu\epsilon\bar\rho_1\right),
\nn\\
&\delta\psi_1=\epsilon\Big(\gamma^\mu  {\cal D}_\mu\phi_1
-\frac{2\pi}k\phi_1|\phi_2|^2+\mu\phi_1\Big),\quad
\delta\psi_2=\epsilon\Big(\gamma^\mu  {\cal D}_\mu\phi_2
-\frac{2\pi}k\phi_2|\phi_1|^2+\mu\phi_2\Big),\nn\\
&\delta\chi_1=-\bar\epsilon\Big(\gamma^\mu {\cal D}_\mu\rho_1
+\frac{2\pi}k\rho_1|\rho_2|^2-\mu\rho_1\Big),\quad\delta\chi_2
=-\bar\epsilon\Big(\gamma^\mu {\cal D}_\mu\rho_2+\frac{2\pi}k\rho_2|\rho_1|^2
-\mu\rho_2\Big).
\end{align}
Unlike the ${\cal N}=4$ Abelian theory, in the current case the set of fields $(\phi_a,\psi_a, a^\mu_a)$ do not mix with the other set $(\rho_a,\chi_a,b^\mu_a)$, and the parts of the Lagrangian that depend on the first and second sets are separately invariant under the ${\cal N}=2$ supersymmetry transformation.
A model similar to Eq. \eqref{N=2CS-2} was considered in Ref. \cite{XS} to describe the phase transitions of quantum antiferromagnets in two spatial dimensions.
Identifying the set of fields ($\phi_1, \rho_1,\cdots)$
with the second set $(\phi_2, \rho_2,\cdots)$, we obtain the well known 2+1-dimensional
${\cal N}=2$ CS matter theory~\cite{Hong:1990yh,Lee:1990it}.

The merits of the ${\cal N}=2$ Abelian theory discussed in this subsection over the ${\cal N}=4$ Abelian theory of Sec. \ref{N=4abelian} are twofold. First, the Abelian projection ansatz in Eq. \eqref{abeans2} is a fluctuation on a supersymmetric vacuum, and it is natural to expect to find the gravity dual. This is true because for a particular supersymmetric vacuum specified by a set of occupation numbers, $(N_n,N_n')$, the dual gravity solution is identified as the $\mathbb Z_k$ quotient of the LLM geometry represented by a droplet with the same occupation numbers \cite{Cheon:2011gv} [see Fig. 1(a)]. The ${\cal N}=2$ Ableian theory is then expected to be dual to a gravity theory on such a background. Second---unlike the case of the ${\cal N}=4$ theory---now the projection ansatz is based on both the $n\times (n+1)$ and the $(n+1)\times n$ matrices. Therefore we can adjust the number of each of these block matrices in such a way that they are not large enough to violate the upper bound set by $k$. In Sec. 4, we will discuss one particular example to clarify this point.

\section{A special vacuum and weakly curved dual geometry}

In Ref. \cite{Polchinski:2000uf}, Polchinski and Strassler discussed
the duality relation between the ${\cal N}=1^*$ theory---a mass-deformed theory of the ${\cal N}=4$ super Yang-Mills
theory---and the type IIB string theory (or IIB supergravity) in the presence of a self dual 5-form flux.
The duality includes quantitative maps for perturbative fluctuations and nonperturbative solitonic objects.

Like the vacua of mABJM theory, the vacua of ${\cal N}=1^*$ theory are discrete due to the mass deformation~\cite{Vafa:1994tf}. This implies that the duality relation of the ${\cal N}=1^*$ theory can be compared with that of the mABJM theory.
As we mentioned previously in Sec. \ref{difficulty},
it was found in Ref. \cite{Cheon:2011gv} that there is
a one-to-one correspondence between the supersymmetric vacuum
space of the mABJM theory and the $\mathbb Z_k$ quotient
of the LLM geometry. It is naturally expected that the mABJM theory is related by duality with
the M-theory on the LLM geometries.
Though this duality relation has not yet been confirmed,
the correspondence between the spectrum of BPS
charged particles and the energy of the fractional M2-branes in
the LLM geometries was obtained in a reliable region~\cite{Cheon:2011gv}.

In the previous section, we constructed the ${\cal N}=2$ Abelian theory on the supersymmetric vacua of the mABJM theory, which is a consistent subset of the mABJM theory.
According to the argument stated in the above paragraph,
one can consider the duality relation for the ${\cal N}=2$ Abelian
theory as some subset in the M-theory on the LLM geometries.
One can consider the correspondence for perturbed fluctuations or nonperturbed solitonic
objects on both sides.

In general, the LLM geometries
include highly curved regions, even in the large-$N$ M2-branes limit.
Gravity theories built on such background geometries should include higher-derivative corrections.
Therefore, in these cases, if one attempts
to find the duality between the field theory fluctuations on some vacua and the gravity fluctuations on the
LLM geometries, then the simple Einstein-Hilbert gravity approximation is not enough.
If one has some criterion for the weakly curved LLM geometries in
the large-$N$ limit, it will be useful in the study of the gauge/gravity duality.
In the Young diagram for $k=1$ shown in Fig. 1(b), small-curvature LLM geometries are only
those where the Young diagram has a small number of corners and a large number
of boxes~\cite{Lin:2004nb}.

In this section, we confirm the above criterion clearly.
We select a special supersymmetric vacuum on the field theory side and
analyze the corresponding dual geometry on the gravity side.
We find that the corresponding dual geometry is weakly curved everywhere
in the large-$N$ limit.

\subsection{A special vacuum in the mABJM theory}\label{specialN2}

Let us consider a particular supersymmetric vacuum of the mABJM theory satisfying the constraints
(2.10) -- (2.11).
Our choice of the occupation numbers is
\begin{align}\label{v0}
N_0=N_1=\cdots = N_p=k,\qquad N_0'=N_1'=\cdots = N_p'=k,
\end{align}
where $p$ represents the number of the ${\cal M}^{(n)}_a$ or $\bar {\cal M}^{(n)}_a$ blocks used to build the vacuum solution.
In the droplet picture for general $k$~\cite{Cheon:2011gv}, the choice \eqref{v0}
represents a droplet with a black strip of length $pk$ above the the Fermi level and a white strip of the same length below the Fermi level, shown in Fig.~1(a).

Here we focus on the case of $k=1$.
Inserting Eq. \eqref{v0} into Eq. (2.10), we obtain the
relation between $p$ and $N$,
\begin{align}
\sum_{n=0}^{p-1}(2n+1)=N.
\end{align}
From this relation we read the number of nonvanishing blocks in the vacuum solutions
of $Z^a$ or $W^{\dagger a}$ in Eq. \eqref{ZW-vacua},
\begin{align}
p= \sqrt{N}.
\end{align}
\begin{figure}
\centerline{\epsfig{figure=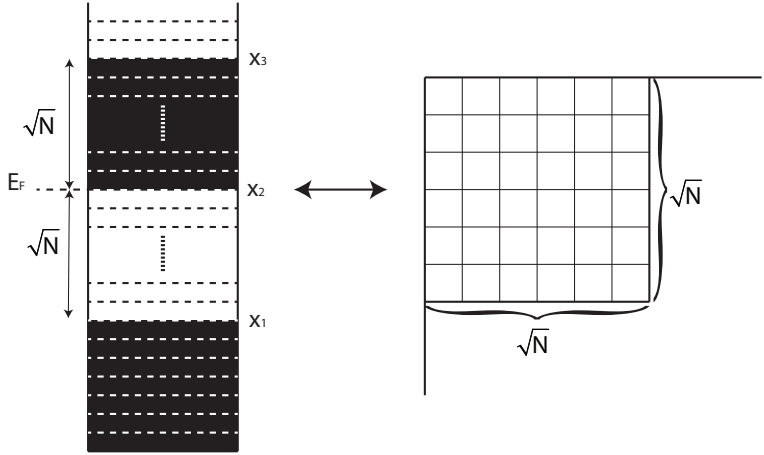,height=60mm}}
\caption{
\small The droplet representation and the corresponding Young diagram for our special vacuum with $k=1$. }
\label{Fig2}
\end{figure}
The droplet and Young diagram for our choice is as in Fig.~2.
If we use this vacuum in the construction of the ${\cal N}=2$ Abelian action \eqref{N=2CS-2},
the overall factors $\beta$ and $\beta'$ are given by
\begin{align}\label{bbp}
\beta=\beta' = \frac12 {\rm tr}\left(S^aS_a^\dagger\right)
=\frac12\sum_{n=1}^{\sqrt{N}-1}n(n+1)
= \frac16\left(N^{3/2} - N^{1/2}\right).
\end{align}
In the next subsection we discuss the dual LLM geometry for the special vacuum \eqref{v0}.

\subsection{Weakly curved dual LLM geometry}

The metric of the LLM geometry is completely determined by two functions $z(x,y)$ and $V(x,y)$, where $(x,y)$ are two of the eight coordinates transverse to the M2-branes. In the notation of Refs. \cite{Lin:2004nb,Cheon:2011gv},
the metric is given by
\begin{align}\label{LLMgeom}
&ds^2 = e^{\frac{4\Phi}{3}}\left(-dt^2 + dw_1^2 + dw_2^2\right) + e^{-\frac{2\Phi}{3}}\Big[h^2\left(dy^2+dx^2\right)
+ y e^G ds_{S^3}^2 + y e^{-G} ds_{\tilde S^3}^2\Big],
\nn \\
&e^{-2\Phi} =\mu_0^{-2}(h^2-h^{-2} V^2), \quad
h^{-2} = 2 y \cosh G, \quad z= \frac12\tanh G,
\end{align}
where $\mu_0$ is a mass parameter fixed by the transverse 4-form field strength.
The functions $z$ and $V$ are
\begin{align}
z(x,y) = \sum_{i=1}^{2m+1}\frac{(-1)^{i+1}(x-x_i)}{2\sqrt{(x-x_i)^2 + y^2}}, \qquad
V(x,y) = \sum_{i=1}^{2m+1}\frac{(-1)^{i+1}}{2\sqrt{(x-x_i)^2 + y^2}},
\end{align}
where the $x_i$'s represent the locations of the boundary lines between the black and white strips in the droplet representation of Fig. 1(b), and $m$ is the number of black or white strips. The black/white strips in such a droplet representation indicate the $\mp \frac12$ values of the function $z$ along the $y=0$ boundary.
For clarity of presentation and easier numerical evaluations, here we consider the case with
a pair of finite-sized black and white strips.
If we fix the $x=0$ position at the Fermi level of the droplet and denote the length of the black strip by $b$
and that of the white strip by $w$, then the functions $z$ and $V$ are given by
\begin{align}\label{weakcurv}
z(x,y) &= \frac12\bigg[\frac{x+b}{\sqrt{(x+b)^2 + y^2}}-\frac{x+b-w}{\sqrt{(x+b-w)^2 + y^2}}+\frac{x-w}{\sqrt{(x-w)^2 + y^2}}\bigg], \nn\\
V(x,y) &= \frac12\bigg[\frac{1}{\sqrt{(x+b)^2 + y^2}}-\frac{1}{\sqrt{(x+b-w)^2 + y^2}}+\frac{1}{\sqrt{(x-w)^2 + y^2}}\bigg].
\end{align}
The Young diagram corresponding to such a geometry is a rectangle with a horizontal side $w$ and vertical side
$b$. Since the number of boxes or the area of the Young diagram is $N$,
we can write $w=N/b$, so that we have only two parameters.

In order to identify the choices of the parameter $b$ that result in a weakly curved geometry in the large-$N$ limit, one has to calculate the Ricci scalar for the metric described by Eq. \eqref{weakcurv}. For a generic spacetime point ($x,y$),
an analytic calculation of the curvature gives a long expression and it is not easy to study.
Therefore, we split the transverse spacetime into three regions according to the distance $r$
from the position of the boundary between black and white strips.
These are the near-boundary region ($r\ll \sqrt N$),
the intermediate region ($r \sim \sqrt N$), and the asymptotic region ($r\gg \sqrt N$).
For the near-boundary and asymptotic regions, analytic treatments are possible for some choices of
the parameter $b$. However, for the intermediate region we can study the geometry only numerically.

By the statement of guage/gravity duality, we expect to have a weak curvature in the large-$N$ limit.
In order to find some LLM geometries that can be the background geometries
in the gauge/gravity duality, we investigate these geometries by evaluating
the Ricci scalar in terms of $b$ and $N$.
We find that the absolute value of the Ricci scalar becomes small at large values of $N$
for some choices of the parameter $b$, while it remains large for the other choices.
In Sec. \ref{difficulty}, we cited that the Young diagram for a weakly curved geometry should have long edges and few corners \cite{Lin:2004nb}. In particular, in the case of Eq. \eqref{weakcurv}, the geometry should be weakly curved if $b=w=\sqrt N$. This is the choice which is dual to the special vacuum used
in Sec. 4.1 (see Fig. 2) for the corresponding droplet and Young diagram.
On the other hand, if $b=1$ and $w=N$, which is the choice of Ref. \cite{Mohammed:2012gi},
the geometry may have some highly curved region.
Since the other choices are in between these two extreme cases, we show that
the former choice corresponds to a weakly curved geometry, while the latter has a highly curved region
in the large-$N$ limit.

The Ricci scalar of the metric described by Eq. \eqref{weakcurv} is a function
of the coordinates $(x,y)$ with the parameters $b~{\rm and}~N$.
\begin{figure}
\centerline{\epsfig{figure=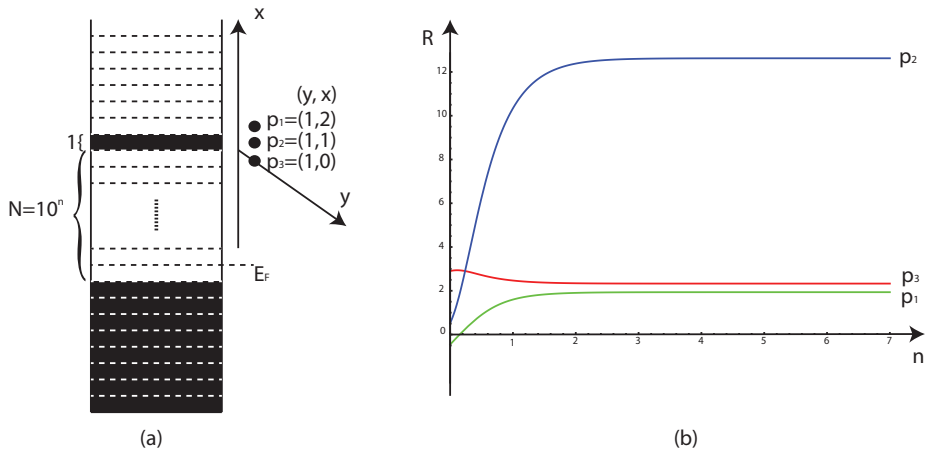,height=60mm}}
\caption{
\small (a) The droplet representation of the extremely asymmetric choice with $b=1$. (b) The plots of the Ricci scalars ($l_{{\rm P}}=1$)
for the extremely asymmetric choice at three selected points in the $(x,y)$ plane.
}
\label{Fig5}
\end{figure}
Before we proceed to the discussion of our choice,
we briefly discuss the extremely asymmetric case with $b=1$ and $w=N$.
We start by plotting the graphs of the Ricci scalar against $N$ at some selected points.
The plots in Fig. 3 show that the Ricci scalars for the selected points
approach significantly large constant values, as we increase $N$.
Though we only take several points into account in these plots, the numerical results are enough to conclude that the extremely asymmetric choices ($b\ll w$ or $w\ll b$)
are related to geometries which have some highly curved regions in the large-$N$ limit.
From now on we mainly focus on the symmetric choice with $b=\sqrt N$.

\subsubsection{The near-boundary region ($r\ll \sqrt N$ )}\label{nearbd}

When we calculate the Ricci scalar in this region, analytic approaches using the large-$N$ expansion are possible.\footnote{Since $r/\sqrt{N}$ is the expansion parameter of the LLM metric near the boundary region, the large-$N$ expansion is the same as the small-$r$ expansion.}
Since the behavior of the geometry near the boundaries at $x_2$ and $x_1$ (or $x_3$) in Fig. 2 are
slightly different, we investigate these two cases separately.
\\

\noindent
(a) $x=x_2$ region:
We call this boundary the symmetric boundary.
To obtain the behavior of the metric \eqref{LLMgeom}, it is convenient to use polar coordinates
$(\tilde r,\theta)$,
\begin{align}\label{polarc}
x= \tilde r \cos\theta,\qquad y = \tilde r\sin\theta,\qquad (0\le\theta\le\pi),
\end{align}
where $\tilde r$ represents the distance from the boundary.
The metric behavior near this boundary is given by
\begin{align}\label{nearF}
ds^2&=\mu_0^{\frac43}\Big(\frac{\tilde N}{4}\Big)^{\frac13}
\Big(1+\frac{2\tilde r}{3\sqrt{\tilde N}}\Big)ds_{{\rm R}^{2+1}}^2
+\frac{2\mu_0^{-\frac23}}{\tilde r}\Big(\frac{4}{\tilde N}\Big)^{\frac16}
\Big(1-\frac{\tilde r}{3\sqrt{\tilde N}}\Big)\Big(\frac14 d\tilde r^2+\tilde r^2ds_{S^7}^2\Big)
\nn \\
&= \pi^{\frac23}(l_{\rm P}\mu_0)^2  N^{\frac13}\bigg[\Big(1 + \frac{2 r}{3 {\sqrt N}}\Big)
ds_{{\rm R}^{2+1}}^2
+\frac4{\mu_0^2 r\sqrt N}\Big(1 - \frac{ r}{3 {\sqrt N}}\Big)\Big(\frac14 d r^2+r^2ds_{S^7}^2\Big)\bigg],
\end{align}
where $ds_{{\rm R}^{2+1}}^2$ represents the worldvolume metric of the M2-branes.
In the second line of Eq. \eqref{nearF} we have used
\begin{align}\label{rN}
\tilde r=2\pi l_{\rm P}^3\mu_0\, r,\qquad \tilde N =
(2\pi l_{\rm P}^3\mu_0)^2 N,
\end{align}
where $r$ is the rescaled dimensionless coordinate.
The rescaling in Eq. \eqref{rN} is based on the flux quantization in 11-dimensional supergravity.
The quantization implies
\begin{align}
x_{i+1} - x_i = 2\pi l_p^3 \mu_0 \mathbb{Z}.
\end{align}

At the boundary $(r=0)$, the metric \eqref{nearF} has a coordinate singularity.
To clearly see the behavior of the geometry at the boundary, we consider
the coordinate transformation
\begin{align}\label{rrho}
  r=\rho^2.
\end{align}
Then the metric \eqref{nearF} is rewritten as
\begin{align}
ds^2
= \pi^{\frac23}(l_{\rm P}\mu_0)^2  N^{\frac13}\bigg[\Big(1 + \frac{2 \rho^2}{3 {\sqrt N}}\Big)
ds_{{\rm R}^{2+1}}^2
+\frac4{\mu_0^2 \sqrt N}\Big(1 - \frac{ \rho^2}{3 {\sqrt N}}\Big)\Big( d \rho^2+\rho^2ds_{S^7}^2\Big)\bigg].
\end{align}
At $\rho=0$, we have the metric of ${\mathbb R}^8$ along the space transverse to the
M2-branes~\cite{Cheon:2011gv}.
Near the symmetric boundary the Ricci scalar has the form
\begin{align}\label{FermiR}
{\cal R}=\frac4{3\pi^{\frac23}l_{\rm P}^2}\Big(\frac{1 }{\sqrt N}\Big)^{\frac23}\bigg[1+\frac{29}{48}\frac{ \rho^2}{\sqrt N}
+{\cal O}\Big(\frac{ \rho^2}{\sqrt N}\Big)^{2}\bigg].
\end{align}
This shows that the Ricci scalar is positive and decreasing with increasing $N$.
We also notice that the Ricci scalar has a local minimum at the boundary; see Fig. 5.
\begin{figure}
\centerline{\epsfig{figure=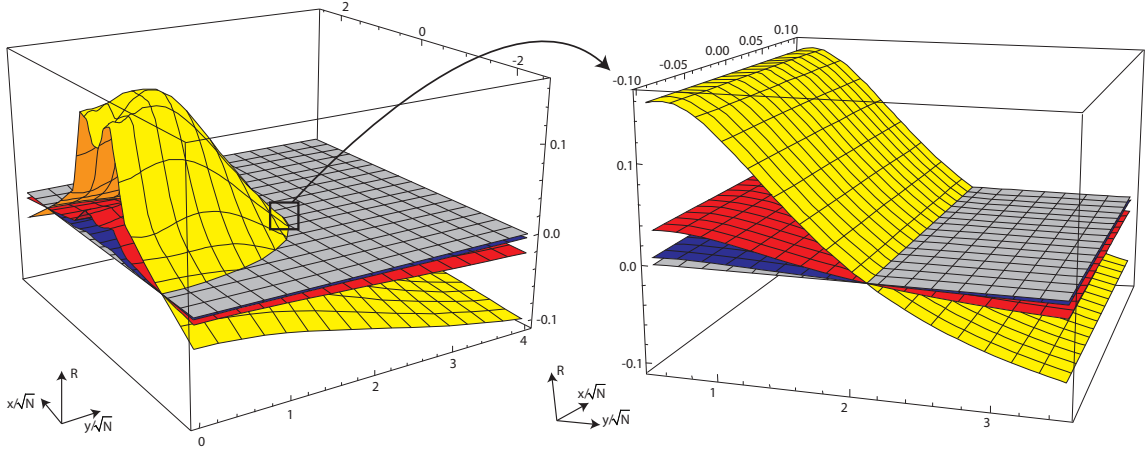,height=60mm}}
\caption{
\small Three-dimensional plots of the Ricci scalars versus the $(x,y)$ coordinates for $\sqrt N=10^2$ (yellow),  $\sqrt N=10^3$ (red), and $\sqrt N=10^4$ (blue). The $(x,y)$ plane in gray is included to trace the line of the zero Ricci scalar.
}
\label{Fig4}
\end{figure}

The regularity of the geometry can also be confirmed by evaluating other invariants.
For instance, the contraction of the Ricci tensor ${\cal R}_{MN}$ is given by
\begin{align}\label{Ricci}
{\cal R}_{MN}{\cal R}^{MN}=\frac{80}{9\pi^{\frac43}l_{\rm P}^4}\Big(\frac1N\Big)^{\frac23}\Big[1+\frac{371}{120}\frac{\rho^2}{\sqrt N}+{\cal O}\Big(\frac{\rho^2}{\sqrt N}\Big)^2\Big].
\end{align}
Similarly, the evaluation of the Kretschmann invariant gives the following regular result
\begin{align}\label{Kret}
{\cal R}_{KLMN}{\cal R}^{KLMN}=\frac{52}{9\pi^{\frac43}l_{\rm P}^4}\Big(\frac1N\Big)^{\frac23}\Big[1+\frac{449}{156}\frac{\rho^2}{\sqrt N}+{\cal O}\Big(\frac{\rho^2}{\sqrt N}\Big)^2\Big],
\end{align}
where ${\cal R}_{LKMN}$ is the Reimann tensor.
The expressions \eqref{Ricci} and \eqref{Kret} also confirm that
our geometry is weakly curved in the large-$N$ limit.
This analysis can be repeated for other regions.
\\
\noindent
(b) $x=x_1$ region:
Due to the symmetry of our droplet, the Ricci scalar near the boundary at $x=x_3$
is the same as the one near the boundary at $x=x_1$.
Using the polar coordinates in Eq. \eqref{polarc} with $\tilde r$ indicating the distance
from the boundary at $x_1$, we obtain the metric near this boundary,
\begin{align}\label{otherb}
ds^2= &(4\pi)^{\frac23}(l_{\rm P}\mu_0)^2  N^{\frac13}\nn\\
&\times\bigg[\Big(1 + \frac{(1-9\cos\theta )r}{6 {\sqrt N}}\Big)
ds_{{\rm R}^{2+1}}^2
+\frac1{\mu_0^2 r\sqrt N}\Big(1 - \frac{(1-9\cos\theta ) r}{12 {\sqrt N}}\Big)\Big(\frac14 d r^2+r^2ds_{S^7}^2\Big)\bigg].
\end{align}

We use the coordinate transformation \eqref{rrho} to see the behavior of the geometry
at $r=0$. We obtain the metric of ${\mathbb R}^8$, and the corresponding
Ricci scalar is given by
\begin{align}\label{otherbR}
{\cal R}=\frac{2^{\frac23}}{3\pi^{\frac23}l_{\rm P}^2}\Big(\frac{1 }{\sqrt N}\Big)^{\frac23}\bigg[&1-\frac{107-45\cos\theta}{48}\frac{\rho^2}{\sqrt N}
+{\cal O}\Big(\frac{ \rho^2}{\sqrt N}\Big)^{2}\bigg].
\end{align}
This also shows that the Ricci scalar is decreasing with increasing $N$.
It has a local maximum at the boundary ($\rho=0$); see Fig. 5.

\begin{figure}
\centerline{\epsfig{figure=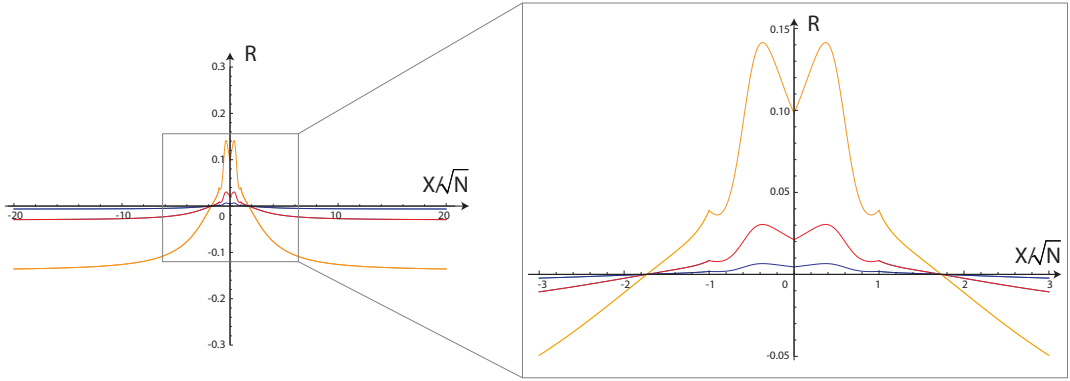,height=50mm}}
\caption{
\small The plot of the Ricci scalar versus the $x$ coordinate at $y=0$ for $\sqrt N=10^2$ (yellow),  $\sqrt N=10^3$ (red), and $\sqrt N=10^4$ (blue).
}
\label{Fig51}
\end{figure}
\subsubsection{The intermediate region ($r \sim \sqrt N$)}
As we stated above, for the intermediate region we can only rely on numerical methods to study the curvature.
In the case of our symmetric choice, the absolute value of the curvature is decreasing
in the entire intermediate region with increasing $N$.
To verify this fact, we plot some three-dimensional graphs of the Ricci scalars versus the coordinates $(x,y)$
for large values of $N$; see Fig. 4.
In this three-dimensional plot, we exclude the numerically problematic regions near $y=0$, which need special treatments.
In the regions where the numerics behave well, the three-dimensional plot in Fig. 4 shows that the geometry
is weakly curved in the large-$N$ limit. The plot also shows that the Ricci scalar is positive when the distance $r$ from the symmetric boundary is small, and it is negative when this distance is large. In between these two regions there is an ellipse-like line in the $(x,y)$ plane where the Ricci scalar is zero.

The region near the $y=0$ boundary can be treated in either of the following two ways. If the distance from any of the boundaries separating the black and white strips is much less than $\sqrt N$, then this region falls under the same category as the near-boundary region, which has already been discussed in Sec. \ref{nearbd}. On the other hand, if this distance is of the order of $\sqrt N$, we expand the functions $z$ and $V$ in powers of $y$ to the leading orders, and then calculate the Ricci scalar from these expansions. The plot of such simplified Ricci scalars versus $N$ also confirms the claim that the curvature is small in the intermediate regions; see Fig. 5. The plots also show that the Ricci scalar is decreasing with increasing $N$.

\subsubsection{The asymptotic region ($r\gg \sqrt N$)}
Like the near-boundary limit ($r\ll \sqrt N$), for the asymptotic limit $(r\gg \sqrt N)$
an analytic treatment is possible.
We proceed by expanding the function $z$ and $V$ for large $r$
as\footnote{Here $r$ represents
the distance from the symmetric boundary.}
\begin{align}\label{zVasymp}
&z(r,\theta) \approx \frac12\cos\theta - \frac{\sin^2\theta}{2 r}\sum_{i=1}^{2m+1} (-1)^{i+1} x_i
- \frac{3\cos\theta\sin^2\theta}{4 r^2}\sum_{i=1}^{2m+1} (-1)^{i+1} x_i^2,
\nn \\
&V(r,\theta)\approx \frac1{2r} + \frac{\cos\theta}{2 r^2}\sum_{i=1}^{2m+1} (-1)^{i+1} x_i
+ \frac{3\cos^2\theta -1}{4 r^3}\sum_{i=1}^{2m+1} (-1)^{i+1} x_i^2.
\end{align}
Since the detailed shapes of droplets are not distinguishable in the asymptotic limit,
the metric only depends on $N$.
For this reason, we here treat an arbitrary droplet.
With the above expansions we obtain
\begin{align}\label{asympfuns}
e^{-2\Phi}\approx \frac{\tilde N}{\mu_0^2r^3},\quad
h^2\approx \frac1{2r},\quad
e^G\approx \cot\frac{\theta}{2},
\end{align}
where the parameter $\tilde N$ is given by
\begin{align}
\tilde N = \frac12\left(\sum_{i=1}^{2m+1} (-1)^{i+1} x_i^2
-\sum_{i=1}^{2m+1}\sum_{j=1}^{2m+1} (-1)^{i+j} x_i x_j\right).
\end{align}
The relation between $\tilde N$ and $N$ was given in Eq. \eqref{rN}.

Inserting Eq. \eqref{asympfuns} into Eq. \eqref{LLMgeom}, we obtain
the asymptotic metric
\begin{align}\label{asympmet}
ds_{{\rm asymp}}^2&= \frac{R^2}{4}\left[\left(\frac{8\pi l_{\rm P}^3\mu_0\tilde r}{R^3}\right)^2\left(-dt^2
+ dw_1^2 + dw_2^2\right) + \frac{d\tilde r^2}{\tilde r^2}\right] + R^2 ds_{{\rm S}^7}^2
\nn \\
&= \left(\frac{R}{2}\right)^2 ds_{{\rm AdS}_4}^2 + R^2 ds_{{\rm S}^7}^2,
\end{align}
where $R= (2^5\pi^2 N)^{1/6}l_{\rm P}$.
Since $\tilde r$ has a rescaling symmetry in the first line of Eq. \eqref{asympmet},
one can see that the asymptotic AdS$_4\times$S$^7$ geometry does not depend on the mass parameter
$\mu_0$. We also notice that the result is independent of the shape of the droplet. Therefore, for any droplet with $N$ M2-branes---including our special choice---the asymptotic geometry is AdS$_4\times {\rm S}^7$ and it is weakly curved in the large-$N$ limit, as expected. This completes the verification of the fact that in the case of our symmetric choice the LLM geometry is weakly curved everywhere in the large-$N$ limit.

\section{Conclusion}
In this paper we have constructed ${\cal N}=2,4$ supersymmetric Abelian projections of the mABJM theory. We selected an ${\cal N}=2$ supersymmetric Abelian theory on a particular vacuum of the mABJM theory
and verified that the background geometry of its gravity dual is weakly curved everywhere. Our work was motivated by a potential application of the Abelian projected mABJM theory to describe some condensed matter systems. If an effective action of a condensed matter system is formulated as a truncation of well-established theory
with a well-understood dual gravity,
the strong-coupling regime of the system can be studied via gauge/gravity duality. In order for this to be realized, the background geometry of the gravity dual must be weakly curved.

Our first attempt in identifying such a theory was based on a consistent truncation of the mABJM theory with ${\cal N}=4$ supersymmetry.
We found such a truncation problematic because of either of the following two reasons.
First, some of the truncation ans\"atze violate the conditions required by the supersymmetry invariance of the vacua at the quantum level. Since the gauge/gravity duality maps the supersymmetric vacuum solutions of the mABJM theory to the ${\mathbb Z}_k$ quotient of LLM geometry, the map is unclear for the Abelian theories built on nonsupersymmetric vacua. Second, when the truncation ans\"atze do not violate the condition required by quantum supersymmetric vacua, we found that the dual geometry has highly curved regions, and the gravity approximation of the fluctuations on such a geometry cannot be trusted.

In order to overcome these problems, we constructed an ${\cal N}=2$ Abelian theory by using another consistent truncation of the mABJM theory. Our truncation ans\"atze involved special fluctuations on the supersymmetric vacua of the mABJM theory, and the gravity duals can be built as fluctuations on the geometries that are dual to these supersymmetric vacua. It turns out that some of the dual geometries contain highly curved regions and the study of gravity theories on those geometries should include higher-derivative corrections. For this reason, focusing on the $k=1$ case, we selected a particular vacuum for which the dual LLM geometry is weakly curved everywhere. We carried out a detailed study of this geometry.

Our assessment can be repeated for a more general LLM geometry, but in this paper we focused on the geometries for which the droplet representation has only a single pair of finite-length black and white strips. The metric is characterized by three parameters, which are the mass parameter, the lengths of the black/white strips, and the total number of M2-branes ($N$).
We found that the geometry is weakly curved when the lengths of the black and white strips are the same and equal to $\sqrt N$ in the large-$N$ limit. To verify this, we studied the behavior of the metric by splitting the transverse space into three intervals depending on the distance $r$ from the boundary between the black and white strips.
For the near-boundary region ($r\ll \sqrt N$) and the asymptotic region ($r\gg \sqrt N$), we showed analytically that the absolute value of the Ricci scalar is small and decreases with increasing $N$. We found that the Ricci scalar is always positive near the boundary whereas it is negative in the asymptotic region, in agreement with the expectation that the geometry is AdS$_4\times {\rm S}_7$ in the latter region. In the intermediate region ($r \sim \sqrt N$), we could study the geometry only numerically. By plotting the graphs of the Ricci scalar versus the transverse space coordinates, we verified that in the intermediate regions the geometry is weakly curved as well, and the curvature decreases with increasing $N$.

In this paper we clarified the map between the vacuum of a special ${\cal N}=2$ Abelian projected mABJM theory and the LLM background geometry of the gravity dual. It still remains to figure out the type of fluctuations on this geometry that result in the gravity theory dual to the ${\cal N}=2$ Abelian theory. There are also other interesting aspects of the map. It is evident that the ${\cal N}=2$ Abelian theory supports soliton solutions, such as vortices and domain walls. It is interesting to find the corresponding objects in the gravity dual. One can also consider more general fluctuations on our special vacuum and study the gravity dual. These issues will be discussed elsewhere.

\section*{Acknowledgements}
We would like to thank Min-Young Choi, Shinsuke Kawai, Hee Cheol Kim,
Kyung Kiu Kim, Seok Kim, Chanyong Park, and Corneliu Sochichiu for helpful discussions.
We also thank Hee Cheol Kim for comments on the behavior of the Ricci scalar
at the boundaries of droplets.
This work was supported by the Korea Research Foundation Grant
funded by the Korean Government with grant numbers 2011-0011660 (Y.K.), 2011-0009972 (O.K.),
and by the World Class University grant no. R32-10130 (O.K.).

\end{document}